\documentclass[aps,prd,twocolumn,groupedaddress,hypertex]{revtex4-1}
 
\usepackage{amsmath}
\usepackage{amssymb}
\usepackage{verbatim}
\usepackage{url}
\usepackage{hyperref}

\begin{document}

\title{A Constraint-Based Approach to the Chiral Magnetic Effect}
\author{Lionel Brits}
\email{lbrits@phas.ubc.ca}
\author{James Charbonneau}
\email{james@phas.ubc.ca}
\affiliation{Department of Physics \& Astronomy, University of British Columbia, \\
6224 Agricultural Road, Vancouver, B.C. V6T 1Z1, Canada }

\begin{abstract}
We propose a way to introduce the currents responsible for the chiral magnetic effect, and similar phenomena, into the AdS/CFT description.  Such currents are thought to occur in heavy ion collisions due to topologically nontrivial field configurations and in dense stars due to beta decay. They may be responsible for the $\mathcal{P}$ and $\mathcal{CP}$ odd effects seen at RHIC and the anomalously large velocities observed in some pulsars.  We discuss the boundary conditions that allow the phenomenon to exist in real systems and show how one would introduce similar boundary conditions into a holographic model of QCD such that the current is reproduced. 
\end{abstract}

\maketitle

\section{Introduction}
Recently much attention has been given to the effects of currents that arise from the axial anomaly in QCD. Such currents  arise in the presence of a magnetic field when there is an imbalance in the number of right- and left-handed fermions. The most well known of these are the charge separation effect \cite{Kharzeev:2007tn} and the chiral magnetic effect \cite{Fukushima:2008xe}. There is a large body of literature describing these phenomena that begins with the study of topological currents in condensed matter systems \cite{Alekseev:1998ds}, and includes anomalous axion interactions in QCD \cite{Metlitski:2005pr} and the high density analogue of the Chiral Magnetic Effect in dense stars \cite{Charbonneau:2007db,Charbonneau:2009ax}. The motivation for studying these currents is the actual observation of local $\mathcal{P}$ and $\mathcal{CP}$ violation in QCD at RHIC \cite{Voloshin:2004vk,Selyuzhenkov:2005xa,Voloshin:2009hr,Abelev:2009uh,Abelev:2009txa}, and the possibility that they are responsible for generating the large velocities (kicks) seen in some pulsars \cite{Charbonneau:2009hq}. 

While the initial investigations of topological currents were made using field theory, there have also been attempts to find evidence for these currents using string theory, which resulted in a debate over whether or not these string models are complete \cite{Rebhan:2009vc,Gorsky:2010xu}. It was recently pointed out in \cite{Rubakov:2010qi} that the source of the confusion lies in incorporating nonzero values of axial charge into these models. In previous derivations \cite{Yee:2009vw,Rebhan:2009vc} a temporal component of a static axial background field was used to mimic an axial chemical potential. This leads to the eventual cancellation of the current by Bardeen counter-terms. A possible solution to the problem is that a chemical potential can only be introduced to a conserved quantum number, which is facilitated by a redefinition of axial charge \cite{Rubakov:2010qi}.

We propose an alternative viewpoint. The real distinction that must be made is between introducing state variables and background fields. We will see that the solution to this problem is present in all examples of the topological currents  derived from considerations of the axial anomaly. The key is that external constraints exist in all the physical realizations of the current. Straight-forward field theory manipulations alone cannot result in a current. There is always another constraint that must be enforced. Each system has a mechanism that forces the axial charge to be fixed, if not strictly conserved, even though this state may only exist for a short time. 

We will begin with a brief introduction of a simple AdS/QCD model in Section \ref{softwall}, which is the model we will use to illustrate our point. In this section we will solve the equations of motion and derive the expressions for the current using the holography. In Section \ref{section:anomalycancelation} we will discuss how the introduction of Bardeen counter-terms restores the conservation of the vector current. In Section \ref{ADSCFT:fail} we will show how this holographic model manages to reproduce the topological axial current,
\begin{align}
\label{eq:topax}
J^{3}_A =\frac{N_c}{2\pi^2} B \mu\,,
\end{align}
 but fails to reproduce the topological vector current, 
 \begin{align}
 \label{eq:topvec}
J^{3}_V =\frac{N_c}{2\pi^2} B \mu_5\,,
\end{align}
 when using the standard boundary conditions. We will review a possible solution to this problem in Section \ref{section:rubakov} that involves reconsidering what the appropriate boundary conditions are. We will then introduce our solution to the problem in Section \ref{section:constraints}. 

\section{Soft-wall AdS/QCD}
\label{softwall} 
In the AdS/QCD model \cite{Polchinski:2001tt,Erlich:2005qh,DaRold:2005zs}, the bottom-up approach is followed. The holographic dual lives on a 5D AdS space with the background metric
\begin{equation}
ds^2 = \frac{1}{z^2} \left( \eta_{\mu\nu} dx^\mu dx^\nu + dz^2\right),
\end{equation}
with \(\eta_{\mu\nu} = \operatorname{diag}(-1,1,1,1)\). The well-known conformal symmetry of this space reflects the near-conformal behaviour of QCD in the UV (\(z \to 0\) in the dual). To break this symmetry at low energies, a ``hard wall'' is introduced in the form of a brane in the IR (\(z \to \infty\)), effectively cutting off the space. In order to obtain the correct Regge behaviour, the ``soft-wall'' model was introduced by \cite{Andreev:2006vy,Karch:2006pv}, where the space is smoothly cut off by turning on the dilaton,
\begin{equation}
\Phi(z) = z^2.
\end{equation}
In four-dimensions we are interested in calculating currents. These four-dimensional operators have five-dimensional fields as holographic counterparts. We are interested in axial and vector currents, so the five-dimensional fields are left and right-handed gauge fields $L$ and $R$. The na\"ive global \(SU(N_f)_L \times SU(N_f)_R\) flavour symmetry of the field theory becomes a local, gauge symmetry in the bulk, and the associated  Noether currents \(J_L\) and \(J_R\) are dual to gauge fields $L$ and $R$. 
We will focus on the case of a single flavour, as the extension to \(N_f > 1\) is not relevant to our discussion.  The dynamics of the gauge fields  in the bulk are described by the action
\begin{align}
\label{eq:action}
S = S_{YM}[L] + S_{YM}[R]  + S_{CS}[L] - S_{CS}[R],
\end{align}
where the components of the action are made from the quadratic expansion of the Dirac-Born-Infeld (DBI) action and Chern-Simons (CS) terms, 
\begin{align}
S_{YM}[A] &= - \frac{1}{8 g_5^2} \int\! dz\, d^4x\ e^{-\Phi} \sqrt{-g}\,F_{MN}F^{MN} ,
\\S_{CS}[A] &= -\frac{N_c}{24\pi^2} \int\! dz\, d^4x \  \epsilon^{MNQPR} A_M F_{NQ} F_{PR}\,.
\end{align}
The uppercase Latin  indices run over all five coordinates $\{0,1,2,3,z\}$. The standard soft-wall model action has a scalar and pseudoscalar sector that we will ignore. 

Here a Chern-Simons term has been included in order to reproduce the chiral anomaly \cite{Domokos:2007kt}. Note that the Chern-Simons action is not gauge invariant in that its variation under a gauge transformation \(A \to A + d\alpha\) is a surface term. Since \(A\) is coupled to a chiral fermion on the boundary, one may think of this variation as being cancelled by the variation of the boundary action, \(\delta S = - \int\!d^4x\, \alpha \partial_\mu J^\mu \), by virtue of the anomaly \cite{Hill:2006ei}. If we think of the holographic relationship between the bulk and boundary theories as arising from taking some limit of a larger theory (D-branes, for example) then this is the statement that the larger theory is completely anomaly free, but that the anomaly cancellation does not survive the decoupling limit. If we wish to study, for example, the boundary theory in isolation, we must cancel the anomaly by hand (by renormalization). It is not possible to do so for both left- and right-handed transformations (equivalently, vector and axial transformations) by means of local counter-terms, but we can at least cancel the vector anomaly, which is enough to keep the theory consistent. The counter-term, given by Bardeen~\cite{Bardeen:1969md}, and introduced in the holographic setting in \cite{Hill:2006ei,Rebhan:2009vc,Rubakov:2010qi,Gynther:2010ed,Gorsky:2010xu}, takes the form of a boundary term in the 5D bulk action,
\begin{equation}
\label{eq:bardeenaction}
S_\mathrm{Bardeen} = -\frac{N_c}{12\pi^2} \int\!d^4x\  \epsilon^{\mu\nu\rho\sigma} L_\mu R_\nu (L_{\rho\sigma}+R_{\rho\sigma}),
\end{equation}
where $L_{\mu\nu}$ and $R_{\mu\nu}$ denote the left- and right-handed field tensors. This term is introduced to ensure that the vector current is free of local anomalies. We will see in Section \ref{section:boundaryconditions} that the cancellation of the vector current due to these terms is not complete if we allow for sources in the bulk. This source in the bulk introduces a global anomaly responsible for the topological current.

\subsection{Correlation Functions}
\label{section:boundary}
Correlation functions involving the current operators \( J^\mu \sim \overline q \gamma^\mu q \) in the 4D gauge theory are obtained by deforming the action by a small amount  \(\delta S \propto \int\!d^4x\,j_\mu J^\mu\), and formally expanding the partition function order-by-order in \(j_\mu\). In the holographic correspondence such a deformation is dual to placing probes near the boundary in the 5D bulk theory. This amounts to deforming the boundary conditions for the gauge field \(A_a\) by a small amount \(j_a\). Similarly, we can introduce a background field \(B_\mu\) for the quarks by adding \( \overline q \gamma^\mu b_\mu q \) to the Lagrangian of the 4D theory, deforming the boundary condition for \(A_a\) by a further amount \(b_a\) in the bulk. We will therefore require the following behaviour of the fields \(A^L\) and \(A^R\) near the boundary:
\begin{align}
\label{eq:BC0}
A^L_\mu(z\to 0) &\to j^L_\mu + b^L_\mu,\nonumber
\\A^R_\mu(z\to 0) &\to j^R_\mu + b^R_\mu.
\end{align}

The electromagnetic currents are derived by taking derivatives of the action with respect to $j^{L/R}_\mu $ as these are dual to the current
\begin{align}
\label{eq:getcurrent}
\mathcal{J}_\mu^{L/R}  = \frac{\delta S}{\delta j^{L/R}_\mu },
\end{align}
where \(S\) is evaluated over a classical solution. Varying the action yields 
\begin{align}
\delta S &= \int_M\! \left[ \frac{\delta \mathcal L}{\delta A_a} - \partial_b\left( \frac{\delta \mathcal L}{\delta \partial_b A_a} \right) \right] \delta A_a + \int_{\partial M} \! n_b \left( \frac{\delta \mathcal L}{\delta \partial_b A_a}   \delta A_a \right),
\end{align}
where $n_b$ is normal to the boundary. This variation indicates we may evaluate Eqn.~(\ref{eq:getcurrent}) using the equation
\begin{align}
\label{eq:getcurrent2}
\mathcal{J}_\mu^{L/R}  =  \left.\frac{\delta \mathcal{L}}{\delta \partial_z A^{L/R}_\mu }\right|_{z=0} + \dots\,,
\end{align}
where \(\dots\) represents the variation of the boundary contribution to the action. The symmetric combination of the boundary conditions we will use are
\begin{subequations}
\begin{align}
\label{eq:BC1}
b_0^{L} + b_0^{R} &= \mu,
\\b_i^{L}+ b_i^R &= -x_j B_k,
\end{align}
\end{subequations}
where we have chosen $ijk$ to be the even permutations of $123$. 

\subsection{The Equations of Motion}
The first thing we will do is establish the equations of motion. We work in the $A_z=0 $ gauge which yields a Yang-Mills action
\begin{align}
S_{YM}[A] =  & - \frac{1}{4 g_5^2} \int\! dz\, d^4x\ \left\{ \frac{e^{-\phi}}{z} A_\mu(\box \eta^{\mu\nu} - \partial^\mu\partial^\nu)A_\nu  \right. \\ &\left.+ A_\mu \partial_z  \frac{e^{-\phi}}{z} \partial_z A^\mu  \right\}
\\ & + \left. \frac{1}{4 g_5^2} \int\! d^4x\    \frac{e^{-\phi}}{z}A_\mu \partial_z A^\mu \right|_{z=0}^{z=\infty}.
\end{align}
Varying the Yang-Mills part of the action Eqn.~(\ref{eq:action}) with respect to the left-handed potential yields
\begin{align}
\frac{\delta S_{YM}[L] }{\delta L_\mu} = -\frac{R}{2g^2_5}\left[\frac{e^{-\phi}}{z}(\Box \eta^{\mu\nu} - \partial^\mu\partial^\nu)L_\nu + \partial_z\left(\frac{e^{-\phi}}{z} \partial_z L^\mu \right) \right]\,.
\end{align}
Varying with respect to the right-handed potential yields a similar result. The Chern-Simons term yields
\begin{align}
\frac{\delta S_{CS}[L] }{\delta L_\mu} =\frac{kN_c}{2\pi^2}\epsilon^{\mu\nu\rho\sigma}\partial_zL_\nu F_{\rho\sigma}\,.
\end{align}
Using the AdS/CFT relation $\frac{R}{g^2_5} = \frac{N_c}{12\pi^2}$ allows us to write the equations of motion as
\begin{align}
\partial_z \left(  \frac{e^{-\phi}}{z}\partial_z L^\mu  \right) - 24 \epsilon^{\mu\nu\rho\sigma} \partial_z L_\nu \partial_\rho L_\sigma =& 0\,,\\
\partial_z \left(  \frac{e^{-\phi}}{z}\partial_z R^\mu  \right) + 24 \epsilon^{\mu\nu\rho\sigma} \partial_z R_\nu \partial_\rho R_\sigma =& 0\,.
\end{align}

We can also vary the system with respect to $A_z$ and get an equation of motion that will later be used to write down the divergence of the current. This will be used solely to illustrate the role of the Bardeen counter-term in the preservation of the conservation of the vector current. The Yang-Mills portion gives us 
\begin{align}
\frac{\delta S_{YM}[L] }{\delta L_z} =   -\frac{R}{2g^2_5}\frac{e^{\phi}}{z}\partial_z \partial_\mu L^\mu \,.
\end{align}
The contribution from the CS term is
\begin{align}
\frac{\delta S_{CS}[L] }{\delta L_z}  = \frac{N_c}{2\pi^2} \epsilon^{\mu\nu\rho\sigma} \partial_\mu L_\nu \partial_\rho L_\sigma\,.
\end{align}
Rewriting the constants using the dictionary we get the equation of motion,
\begin{align}
\label{eq:A_zeom}
 \frac{N_c}{24 \pi^2}\frac{e^{\phi}}{z}\partial_z \partial_\mu L^\mu  = \frac{N_c}{2\pi^2} \epsilon^{\mu\nu\rho\sigma} \partial_\mu L_\nu \partial_\rho L_\sigma\,.
\end{align}
A similar equation of motion can be found for the right-handed field $R_\mu$ where the right hand side picks up a negative sign. 

\subsubsection{Specific Solutions}
We are interested in a background magnetic field in the $x^3$ direction and choose $L_1(x_2)= R_1(x_2) = -\frac{1}{2} B x_2$ such that they have only $x_2$ dependence.  We can rewrite these fields in terms of the axial and vector fields $A$ and $V$ by using the relations $V = L+R$ and $A = L-R$. We should note that this convention differs from some papers that define the axial relation as $A=R-L$ and can result in sign differences when comparing equations involving axial components. Using our convention we arrive at the equations of motion
\begin{align}
\partial_z \left(  \frac{e^{-\phi}}{z}\partial_z V_0  \right) + 12 B \, \partial_z A_3 &= 0\,,\\
\partial_z \left(  \frac{e^{-\phi}}{z}\partial_z V_3  \right) + 12 B\,  \partial_z A_0  &= 0\,,\\
\partial_z \left(  \frac{e^{-\phi}}{z}\partial_z A_0  \right) + 12 B \, \partial_z V_3 &= 0\,,\\
\partial_z \left(  \frac{e^{-\phi}}{z}\partial_z A_3  \right) + 12 B\,  \partial_z V_0  &= 0\,.
\end{align}
These can be solved by introducing a new coordinate $w$ with
\begin{align}
\partial_w &= \frac{dz}{dw} \partial_z = \frac{e^{-\phi}}{z}\partial_z\,,
\end{align}
so that the appropriate solutions are
\begin{align}
\label{eq:solutions1}
V_0(z) &= v_{00}  + a_{31} ( e^{ -\beta\, w(z) } -1),\\
V_3(z) &= v_{31}  + a_{01} ( e^{ -\beta\, w(z) } -1),\\
\label{eq:solutions3}
A_0(z) &= (a_{00} + a_{01} ) + a_{01} ( e^{ -\beta\, w(z) } -1),\\ 
\label{eq:solutions4}
A_3(z) &= (a_{30} + a_{31} ) + a_{31} ( e^{ -\beta\, w(z) } -1),
\end{align}
where $\beta = 12B$ and \(w(z)\) has the property that \(w(0) = 0\) and \(w(\infty) = \infty\) such that we discard one set of solutions from the beginning. We must still apply appropriate boundary conditions to these solutions. We have grouped the constants in a way that will make the effect of these boundary conditions more transparent later in the paper.

\subsection{Setting Up the Currents}
We will now introduce the standard setup for holographic currents. As discussed earlier the 4-dimensional currents can be found using 
\begin{align}
J^\mu = \frac{\delta S}{\delta A_\mu(0)} = J^\mu_0 + J^\mu_\mathrm{Bardeen} = \left.\frac{\delta \mathcal{L} }{\delta \partial_z A_\mu }\right|_{z \to 0} + J^\mu_\mathrm{Bardeen}\,.
\end{align}
The Bardeen term is chosen to cancel nonconserved components in the vector current. This is the common procedure in quantum field theory as the vector current is the physical current. This process leaves the axial current anomalous. 

The current receives contributions from both $F_{\mu z}F^{\mu z}$ and $S_{CS}$. The currents that arise from this are
\begin{align}
J_{0,L}^\mu &= -\frac{N_c}{24\pi^2} \frac{e^{-\phi}}{z}\partial_z L^\mu  + \frac{N_c}{6\pi^2} \epsilon^{\mu\nu\rho\sigma} L_{\nu} L_{\rho\sigma} \,,\\
J_{0,R}^\mu &= -\frac{N_c}{24\pi^2} \frac{e^{-\phi}}{z}\partial_z R^\mu  - \frac{N_c}{6\pi^2}  \epsilon^{\mu\nu\rho\sigma} R_{\nu} R_{\rho\sigma}\,.
\end{align}
And the Bardeen currents that come from the Bardeen action in Eqn.~(\ref{eq:bardeenaction}) are
\begin{align}
J^\mu_\mathrm{Bardeen,L}& = -\frac{N_c}{6\pi^2} \epsilon^{\mu\nu\rho\sigma} \left( R_\nu \partial_\rho R_\sigma + 2 R_\nu \partial_\rho L_\sigma   -  L_\nu\partial_\rho R_\sigma \right), \\
J^\mu_\mathrm{Bardeen,R}& = +\frac{N_c}{6\pi^2} \epsilon^{\mu\nu\rho\sigma} \left( L_\nu \partial_\rho L_\sigma + 2 L_\nu \partial_\rho R_\sigma   -  R_\nu\partial_\rho L_\sigma \right).
\end{align} 
As discussed earlier we can rewrite currents in terms of the axial and vector fields $A$ and $V$ by using the relations $V = L+R$ and $A = L-R$,
\begin{align}
\label{eq:currentequations1}
J_{0}^\mu &= -\frac{N_c}{24\pi^2} \partial_w V^\mu  + \frac{N_c}{12\pi^2} \epsilon^{\mu\nu\rho\sigma} \left(  V_{\nu} A_{\rho\sigma} +  A_{\nu} V_{\rho\sigma}\right), \\
\label{eq:currentequations2}
J_{0,A}^\mu &= -\frac{N_c}{24\pi^2} \partial_w A^\mu  + \frac{N_c}{12\pi^2} \epsilon^{\mu\nu\rho\sigma}  \left( V_{\nu}V_{\rho\sigma} + A_{\nu}A_{\rho\sigma}  \right), \\
\label{eq:currentequations3}
J_\mathrm{Bardeen}^\mu &= -\frac{N_c}{24\pi^2} \epsilon^{\mu\nu\rho\sigma} \left(   - 4 A_\nu V_{\rho\sigma} +  2 V_\nu A_{\rho\sigma}
 \right) ,\\
 \label{eq:currentequations4}
J_\mathrm{Bardeen,A}^\mu &= -\frac{N_c}{24\pi^2} \epsilon^{\mu\nu\rho\sigma} 2 V_\nu V_{\rho\sigma}\,.
\end{align}
We have used $V_{\mu\nu}$ and $A_{\mu\nu}$ to denote the field tensors composed of vector and axial fields respectively. 

\section{Cancellation of the Vector Anomaly}
\label{section:anomalycancelation}
The first aspect of the problem that we will consider is the cancellation of the vector anomaly. The Bardeen terms  are introduced to ensure that the vector current is strictly conserved, even in the presence of an axial field.  Consider the divergence of the vector current given by Eqn.~\eqref{eq:currentequations1},
\begin{align}
\partial_\mu J_{0}^\mu &= -\frac{N_c}{24\pi^2} \partial_w \partial_\mu V^\mu  + \frac{N_c}{12\pi^2} \epsilon^{\mu\nu\rho\sigma}\partial_\mu  \left(  V_{\nu} A_{\rho\sigma} +  A_{\nu} V_{\rho\sigma}\right) , \\
& =  - \frac{N_c}{2\pi^2} \epsilon^{\mu\nu\rho\sigma} \partial_\mu V_{\nu} \partial_\rho A_{\sigma} +   \frac{N_c}{3\pi^2} \epsilon^{\mu\nu\rho\sigma}   \left(  \partial_\mu V_{\nu} \partial_\rho A_\sigma \right) , \\
& = -  \frac{N_c}{6\pi^2} \epsilon^{\mu\nu\rho\sigma} \partial_\mu V_{\nu} \partial_\rho A_{\sigma} \,,
\end{align}
where we used the right and left-handed version of Eqn.~\eqref{eq:A_zeom} to write
\begin{align}
 \frac{N_c}{24 \pi^2}\frac{e^{\phi}}{z}\partial_z \partial_\mu V^\mu  = \frac{N_c}{2\pi^2} \epsilon^{\mu\nu\rho\sigma} \partial_\mu V_\nu \partial_\rho A_\sigma\,.
\end{align}
The divergence of the vector Bardeen current given by Eqn.~\eqref{eq:currentequations4} becomes
\begin{align}
\partial_\mu J_\mathrm{Bardeen}^\mu &= -\frac{N_c}{24\pi^2} \epsilon^{\mu\nu\rho\sigma}\partial_\mu  \left(   - 4 A_\nu V_{\rho\sigma} +  2 V_\nu A_{\rho\sigma}   \right), \\
& = \frac{N_c}{6\pi^2} \epsilon^{\mu\nu\rho\sigma} \partial_\mu V_{\nu} \partial_\rho A_{\sigma} \,.
\end{align}
The Bardeen terms also alter the divergence of the axial current. The divergence of the axial current given by Eqn.~\eqref{eq:currentequations2}  is
 \begin{align}
\partial_\mu J_{0,A}^\mu &= -\frac{N_c}{24\pi^2}  \partial_w \partial_\mu A^\mu  + \frac{N_c}{12\pi^2} \epsilon^{\mu\nu\rho\sigma} \partial_\mu \left( V_{\nu}V_{\rho\sigma} + A_{\nu}A_{\rho\sigma}  \right)\,, \\
& = -\frac{N_c}{12\pi^2}\left(  \partial_\mu V_{\nu}\partial_\rho V_{\sigma} + A_{\nu}\partial_\rho A_{\sigma}  \right)\,,
\end{align}
where we used the left and right-handed versions of \eqref{eq:A_zeom} to write
 \begin{align}
 \frac{N_c}{24 \pi^2}\frac{e^{\phi}}{z}\partial_z \partial_\mu A^\mu  = \frac{N_c}{4\pi^2} \epsilon^{\mu\nu\rho\sigma}\left( \partial_\mu V_\nu \partial_\rho V_\sigma + \partial_\mu A_\nu \partial_\rho A_\sigma\right)\,.
\end{align}
The divergence of the axial Bardeen current given by Eqn.~\eqref{eq:currentequations4} is
\begin{align}
\partial_\mu J_\mathrm{Bardeen,A}^\mu &= -\frac{N_c}{6\pi^2} \epsilon^{\mu\nu\rho\sigma} \partial_\mu V_\nu \partial_\rho V_{\sigma}\,.
\end{align}
Without the counter-terms the anomaly appears in both the divergence of the vector and axial currents. This configuration where both currents are anomalous is known as the consistent anomaly. Adding the Bardeen counter-term shifts the anomaly from the vector current to axial current leaving us with the covariant anomaly,
\begin{align}
\partial_\mu J^\mu &= 0\,, \\
\label{eq:covariantanomaly}
\partial_\mu J_A^\mu &= -\frac{N_c}{12\pi^2}\epsilon^{\mu\nu\rho\sigma} \left( 3\partial_\mu V_{\nu}\partial_\rho V_{\sigma}+\partial_\mu A_{\nu}\partial_\rho A_{\sigma}\right) \,. 
\end{align}
The Bardeen terms do their job in conserving the vector current. The difference is that we have implemented the procedure of cancelling the anomaly from the holographic side. 

\section{The Problem}
\label{ADSCFT:fail}
We have defined all the tools necessary to discuss the problem. We want to see how the definition of the axial chemical potential as the temporal component of the axial field fails to reproduce the topological vector current \eqref{eq:topvec} when the Bardeen counter-terms are introduced. The standard technique is to apply the following boundary conditions to the holographic system. 

At $z=\infty$ the boundary conditions are 
\begin{align}
V_0(\infty) &= 0,\\
A_0(\infty) &= 0.
\end{align}
Forcing the vector potential to vanish in the bulk at $z=\infty$ means that 
\begin{align}
\label{eq:vogestozero}
 v_{00} = a_{31} = \mu
\end{align}
as seen in Eqn.~\eqref{eq:solutions1}. This is how the vector chemical potential enters the derivation of the axial current. Forcing the axial field to go to zero in the bulk means that $a_{00} = 0$ in Eqn.~\eqref{eq:solutions3}.

In a related system, the confined phase of the Sakai-Sugimoto model, the condition \(A_\mu(\infty) = 0\) is a natural choice for the following reason. In that model, the two gauge fields live on D-branes that join in the bulk, and thus are actually two branches of a single D-brane. Therefore \(A_\mu(\infty) = 0\) reflects the continuity of the single gauge field. This continuity also reflects the breaking of chiral symmetry in the IR. Note that \(A_\mu(\infty) = 0\) is not a gauge-dependent statement, as the theory is not gauge invariant under axial transformations. It is a statement about continuity.

At \(z=0\) (i.e., the holographic boundary) the vector combination is given by Eqn.~(\ref{eq:BC1}), 
\begin{align}
V_0(0) &= v_{00} =  \mu,\\
V_i(0) &= -x_j B_k \,,
\end{align}
and we let the axial field be,
\begin{align}
A_0(0) &=  a_{01} =  \mu_5,\\
A_i(0) &= 0.
\end{align}
We see that the zeroth component of the axial field is equated with the axial chemical potential. To take the temporal component of the axial field as the axial chemical potential is the common method of introducing and axial chemical potential into the holography. It mirrors the way the vector chemical potential is introduced by equating it with the temporal component of the vector field. By defining the chemical potentials outright we have approached the problem by using the grand canonical ensemble.  

We assume no background axial field strength and take the background magnetic field to be in the $x^3$ direction,
\begin{align}
A_{\mu\nu} &= 0\,, \\
V_{21} &= -B\,,\\ 
V_{12} &= B\,.
\end{align}
We apply these conditions to find the vector current from Eqns.~(\ref{eq:currentequations1}) and (\ref{eq:currentequations3}) and find that it vanishes, 
\begin{align}
J^3 &= J_{0}^3 + J_\mathrm{Bardeen}^3\,,  \\
&  = -\frac{N_c}{24\pi^2} \partial_w V^3  + \frac{3N_c}{12\pi^2} \epsilon^{3\nu\rho\sigma}\, A_{\nu} V_{\rho\sigma}\,, \\
& = \frac{N_c}{2\pi^2} B a_{01} +  \frac{3N_c}{6\pi^2} \epsilon^{3012}  B a_{01}= 0\,,
\end{align}
while the axial current from Eqns.~(\ref{eq:currentequations2}) and (\ref{eq:currentequations4}) gives the result we would expect,
\begin{align}
J_A^3 &= J_{0,A}^3 + J_\mathrm{Bardeen,A}^3\,, \\
& = -\frac{N_c}{24\pi^2} \partial_w A^3  + \frac{N_c}{12\pi^2} \epsilon^{3\nu\rho\sigma} A_{\nu}A_{\rho\sigma} \,, \\
& = \frac{N_c}{2\pi^2}\,B \mu\,,
\end{align}
where we used \eqref{eq:vogestozero} to write this in terms of the chemical potential. 

The contribution from the Bardeen counter-term that was introduced to ensure gauge invariance of the currents has made the contribution to the topological vector current from the axial chemical potential vanish. But, the axial current \eqref{eq:topax} that arises from a vector chemical potential still appears, even in the presence of the counter-terms. This demonstration of the counter-terms cancelling the vector current was originally done using the Sakai-Sugimoto model \cite{Rebhan:2009vc} and lead people to believe that maybe the topological vector current responsible for phenomena like the chiral magnetic effect did not exist in holographic QCD models.  But the effect can be reproduced in many holographic models, like the one we present here. It became apparent that the problem was one of thermodynamics.

\section{Rubakov's Solution}
\label{section:rubakov}
In the previous section, by choosing to set \(V_0(z\to 0)\) equal to the chemical potential, \(\mu\), and \(A_0(z\to 0)\) equal to the axial chemical potential, $\mu_5$ we have chosen to work in the grand canonical ensemble with respect to both the vector fermion number \(N_L + N_R\) and the axial fermion number  \(N_L - N_R\). However, the axial fermion number is not conserved, precisely because of the anomaly (\ref{eq:covariantanomaly}), and so the canonical and grand canonical ensembles do not represent equivalent pictures with regards to this operator. This was pointed out by Rubakov~\cite{Rubakov:2010qi}, among others. Working in the grand canonical ensemble requires us to replace the axial fermion number operator with one that is conserved~\cite{Rubakov:2010qi}. 

Though Rubakov used a different holographic model than the one we employ, it is possible to demonstrate his solution. Rubakov stressed that the temporal component of the axial field is not the same thing as a chemical potential. The chemical potential must be introduced to the action conjugate to a conserved charge. The procedure is to shut off all axial fields $A_\mu$ and define a conserved charge to introduce a chemical potential $\mu_5$. The charge is usually given by the temporal component of the current integrated over all space. With the anomaly present this definition of the charge is not  conserved. To ensure the charge is conserved we use the part responsible for the nonconservation in Eqn.~\eqref{eq:covariantanomaly} to modify the charge,
\begin{align}
\label{eq:rubakovcharge}
Q^5_\text{Rubakov}  = \int\!dx^3 J_0^{5} + \frac{3 N_c}{12\pi^2} \int\!dx^3\epsilon^{ijk}\,  V_{i}\partial_jV_{k}\,.
\end{align}
Note that our axial current is defined as $L-R$ while Rubakov's is defined as $R-L$ giving a sign difference on our anomaly, as well as many of the other equations that follow. This charge is invariant under electromagnetic gauge transformations. We now add a chemical potential to the action,
\begin{align}
S[\mu_5] &= S + \mu_5 \int\! dx^0 Q_\text{Rubakov}^5 \,,\\
\label{eq:modifiedchargecurrent}
& =  \left(S + \mu_5\int\!dx^4 J_0^{5}  \right) + \mu_5\frac{3 N_c}{12\pi^2} \int\!dx^4\epsilon^{ijk}\,  V_{i}\partial_jV_{k}\,,
\end{align}
where $S$ is the original action of the system with the chemical potential introduced in such a way that we can find the current. As shown earlier, the Bardeen counter-terms cause all contributions to the current from $S$ to vanish. The counter-terms also cancel the contributions of the vector field from the axial current $J^5_0$. The only contribution to the current is from the variation of the last term in Eq.~(\ref{eq:modifiedchargecurrent}). The variation produces
\begin{align}
\frac{\delta S}{\delta V_i} =  \mu_5\frac{ N_c}{4\pi^2} \epsilon^{ijk}\,\partial_jV_{k} = \frac{ N_c}{2\pi^2}  \mu_5 B \,,
\end{align}
which is the vector current responsible for the chiral magnetic effect. 

The axial chemical potential takes over the role of the temporal component of the axial field, but without the problems of getting cancelled by the counter-terms. Though this arrives at the correct answer it is unappealing because we have to construct an effective action. The chemical potential is added in such a way it cannot be cancelled and then is used to derive the current. This is a solution, but it would be more desirable to see the current arise from the model itself. This is where our program starts.

\section{A Solution Arising from Boundary Conditions}
\label{section:constraints}
While Rubakov's method reproduced the accepted result, we would like to find a resolution that is contained within the holographic model.  Working entirely in the grand canonical ensemble did not work. Working in the canonical ensemble, however, where one fixes the axial fermion number, is problematic on physical grounds. One imagines a state at some initial time, having some value \(N_L - N_R\), evolving to a later time, and having a different value of \(N_L - N_R\). Therefore, restricting the path integral to states with a fixed value of \(N_L - N_R\) amounts to placing a \textit{constraint} on the system. We will now address whether this constraint is physical.

We will discuss three physical systems in which a constraint fixes the value of $N_L - N_R$: induced topological currents in dense stars \cite{Charbonneau:2009ax}, the charge separation effect \cite{Kharzeev:2007tn} and the chiral magnetic effect  \cite{Fukushima:2008xe}. In each of these systems the constraint is due to the quasistatic nature of the system, particularly, they require a nontrivial boundary condition to induce and maintain an axial charge. For each example we will discuss the external mechanism responsible for introducing the axial charge and maintaining it such that the current can flow. 

The first example is the appearance of topological currents in neutron stars, where the current was derived by considering the microscopic elements of the system  \cite{Alekseev:1998ds,Metlitski:2005pr,Charbonneau:2007db,Charbonneau:2009ax}. Given the complexity of neutron stars it is useful to think about the current in terms of numbers of particles. The current is given by 
\begin{align}
\label{eq:current}
J_\text{V} = (n_l - n_r)\frac{e\Phi}{2\pi}\,,
\end{align}
where $n_r(T,\mu)$ and $n_l(T,\mu)$ are the one-dimensional number densities of left and right-handed Dirac fermions and $\Phi$ is the magnetic flux. This formula tells us that to find the magnitude of the current one has to count the number of particles aligned with the magnetic field and subtract them against those aligned against the magnetic field. In a neutron star the equilibrium processes are given by beta and inverse beta decay, known collectively as the Urca processes. These are weak interactions that act more on left-handed electrons than right-handed electrons. In an infinite system any difference in left and right-handed particles created by the weak interaction would be washed out due to detailed balance; the creation of a left-handed electron is as likely as the time-reversed process, the scattering of an electron and a proton to create a neutron and neutrino. The finite size of the star essentially breaks time-reversal symmetry and allows the electron to escape before it decays through the weak interaction. This causes a current to flow through the bulk of the star. The key is that electrons are constantly being added by beta decay to maintain the difference between left and right-handed electrons. These processes are like a pump that fixes the axial charge. This current may explain the anomalously large pulsar kicks that have been observed \cite{Charbonneau:2009hq}.

The second example is the charge separation effect \cite{Kharzeev:2007tn}, wherein a current appears in regions where there is a dynamical or spatially varying theta angle $\theta(\vec{x},t)$. These regions may occur at RHIC where a collision creates a small bubble where $\theta(\vec{x},t)\neq 0$ within a larger region (e.g., the rest of the world) where $\theta=0$.  The current takes the form
\begin{align}
j_0 = N_c \sum_f \frac{e_f \mu_f}{2N_f \pi^2}\, \vec{\nabla} \theta\cdot \vec{\Omega} \,,\\
j_i = N_c \sum_f \frac{e_f \mu_f}{2N_f \pi^2}\, \partial_0 \theta\, {\Omega}_i \,,
\end{align}
where \(\Omega\) is the angular velocity of the rotating coordinate frame, which mimics a magnetic field. The domain wall created as the theta angle transitions is the reason the current appears and draws charge to the surface of the theta bubble, which we can see if we integrate over the volume of the theta bubble. If the bubble was infinitely large or the gradient of the change was extremely small, the charge would not accumulate on the domain wall.  The finiteness of the bubble has introduced a constraint in the parity violating aspect of the current. When the domain wall disappears, the parity violation in the system disappears and the charge separation vanishes with it. 

The third example is the previously mentioned chiral magnetic effect \cite{Fukushima:2008xe}. The effect is often written as a statement of the current that causes the charge separation, 
\begin{align}
\label{eq:cme}
j_\text{V}= (\mu_l - \mu_r)\frac{e^2B}{(2\pi)^2},
\end{align}
but this current alone does not naturally appear in heavy ion collisions. What is required is an external source to introduce chirality into the system. In heavy ion collisions it may be possible for transitions to occur from one QCD vacuum to another. When this transition occurs, chirality is induced through a change in the QCD winding number $n_w$, which can be written as the charge 
\begin{align}
Q_w =n_w(t =\infty)-n_w(t =-\infty)\,.
\end{align}
If induced during a collision the charge will introduce a difference in the number of left and right-handed particles through the relation
\begin{align}
N_L^f - N_R^f = 2Q_w\,.
\end{align}
The introduction of the topological charge $Q_w$ allows the system to violate parity and induces a current. Only then can the charge separation occur. 

We have seen three examples of how a constraint is used to hold the axial charge fixed during the life of the current. We should note that these are real systems. In particular the charge separation effect and the chiral magnetic effect explain the $\mathcal{P}$ and $\mathcal{CP}$ violation seen at RHIC \cite{Voloshin:2004vk,Selyuzhenkov:2005xa,Voloshin:2009hr,Abelev:2009uh,Abelev:2009txa}. We will not consider the physics behind the constraint, which often is poorly described in a formal field theory setting. We will only assume that one is necessary for the current to be nonzero.

\subsection{New Boundary Conditions}
\label{section:boundaryconditions}

As we have discussed, in previous derivations of the chiral magnetic effect using holography it was common to associate the axial field on the holographic boundary with the chemical potential. In light of the discussion in Section \ref{section:boundary} the current does not appear in a system that is in true thermal equilibrium. As pointed out in \cite{Rubakov:2010qi} assigning the axial chemical potential to $A_0(0)$ is assigning a chemical potential to a quantity that is not conserved. In an effort to find the true conserved quantity we will derive the axial current without assigning a thermodynamic meaning to the boundary value of the axial field. We will then interpret the result afterwards by comparing it to the results found in QCD.  We will find that there is a component of the axial field that survives the anomaly cancelation and that this component must be associated with the induced axial charge of the system. 

By allowing the axial field in the bulk to take nonzero values, $A_0(\infty)\neq0$, we allow the possibility for a source to add new charge into the system and contribute to the current. This is the exact source needed to provide the axial charge with the constraint required to keep it conserved as discussed in Section \ref{section:constraints}.

We now mirror the derivation in Section \ref{ADSCFT:fail}. As before we choose to use the grand canonical ensemble for the vector combination  Eqn.~(\ref{eq:BC1}). At \(z=0\), on the holographic boundary we get
\begin{align}
V_0(0) &= v_{00} =  \mu,\\
V_i(0) &= -x_j B_k.
\end{align}
As discussed we choose not to assign the value of \(A_\mu(0)\) any thermodynamic quantity. In the IR (i.e., \(z=\infty\)) we have more freedom in imposing boundary conditions. 

In our model we wish to impose a nonzero axial charge on the boundary theory. As mentioned above, this is complicated by the fact that the axial charge is not conserved (equivalently, a lack of gauge invariance). Instead we suppose that there is an IR cut-off in the bulk, beyond which some source exists. In a string model like the Sakai-Sugimoto model, such a source would be provided by the endpoints of strings stretching between the flavour branes and D-branes playing the role of instantons. In a bottom-up model, the source must be inserted by hand. The value of \(A_\mu\) at the cut-off depends on the magnitude of this source. Motivated by this observation, we allow for the possibility that $A^\mu(\infty) \neq 0\,$, reflecting the fact that we have a source of axial charge in the far IR. In the hard-wall model one could similarly impose a nonzero charge on the IR boundary by choosing the correct boundary conditions on the brane responsible for the hard IR cutoff. In the hard-wall model this boundary condition is not the result of a source, but is just a characteristic of the model. 

This source beyond the IR represents the external source required to generate the vector current using the axial anomaly in field theory models. As discussed in Section \ref{section:constraints} topological currents need an external source to manifest themselves. In neutron stars this source is the Urca processes constantly producing new left-handed electrons and in the chiral magnetic effect and charge separation effect this source is the topological charge induced by particle collisions. With this source we will be able to induce a topological vector current that survives the Bardeen cancelation.

We should note that this asymmetric treatment of the axial and vector field boundary conditions only arises because we cannot completely remove the anomaly. With the vector field there is no problem with assigning $V_0(0) =\mu$ because there is no anomaly associated with the current and the vector charge is conserved. The chemical potential can be assigned to the boundary value of the vector field. We could take the same approach with the vector field as we are with the axial field, but the boundary condition ends up still being $V_0(0) =\mu$. The special approach for the axial field is required because of the anomaly. If we had instead chosen a different Bardeen counter-term that made the axial current anomaly free and the vector current anomalous the story would be reversed.

By allowing the possibility for $A^\mu(\infty) \neq 0\,$ our boundary conditions at infinity given by the solutions to the equations of motion \eqref{eq:solutions1} and \eqref{eq:solutions3} are
\begin{align}
V_0(\infty) &= 0,\\
A_0(\infty) &= a_{00}.
\end{align}
Forcing the vector current go to zero in the bulk means that $v_{00} = a_{31} = \mu$, which will contribute when we derive the axial current. The vector current is then found by using Eqns. (\ref{eq:currentequations1}) and (\ref{eq:currentequations3}). In our solution we assume no background axial field strength and a background magnetic field in the $x^3$ direction,
\begin{align}
A_{\mu\nu} &= 0\,, \\
V_{21} &= -B\,,\\ 
V_{12} &= B\,.
\end{align}
We apply these to the vector current from Eqns.~(\ref{eq:currentequations1}) and (\ref{eq:currentequations3}), which is further simplified when we choose just the $x_3$ direction for the current along the magnetic field, to get
\begin{align}
J_{0}^3 &= -\frac{N_c}{24\pi^2} \partial_w V^3  + \frac{N_c}{6\pi^2} \epsilon^{3012}   A_{0} B\,.
\end{align}
The Bardeen vector terms are similarly manipulated to give
\begin{align}
J_\mathrm{Bardeen}^3 &= \frac{N_c}{3\pi^2}  \epsilon^{3 0 12} A_0 B .
\end{align}
We now substitute in the solutions given in Eqns.~(\ref{eq:solutions1})--(\ref{eq:solutions4}). To calculate the true current we evaluate this on the boundary at $z=0$. Using $\epsilon^{3012}=-1$ the total current can now be written as
\begin{align}
J^3 &= J^3_0 + J^3_\text{Bardeen}\,.\\
&=\frac{N_c}{2\pi^2} B  (-a_{00}).
\end{align}
Recalling our solution for the axial field, $A_0(\infty) = a_{00}$, we see that not fixing the boundary condition in the bulk has yielded a connection between the UV and the IR. The current on the boundary depends on a boundary condition in the bulk,
\begin{align}
\label{eq:CME}
J^3 =  \frac{N_c}{2\pi^2} B (-A_0(\infty)) \,.
\end{align}
The current vanishes if $A_0(\infty)=0$ and a nonvanishing current depends on a discontinuity in the IR. Maintaining this field in the bulk acts as though a chiral charge is being fixed externally. However, without axial gauge invariance we do not have the equivalent of Gauss' law to relate \(A(\infty)-A(0)\) to the magnitude of the source in the bulk. This would require treating the bulk and boundary in parallel, as discussed in Section \ref{softwall}. In our work we relax the interpretation of the value of the axial field in the bulk, since there is no equivalent Gauss'  law available. 

As a consistency check we can use equations (\ref{eq:currentequations2}) and (\ref{eq:currentequations4}) to find the axial current,
\begin{align}
\label{eq:axialcurrent}
J^{3}_A =\frac{N_c}{2\pi^2} B \mu\,.
\end{align}
This reproduces the standard result for anomalous axial currents first discussed in \cite{Metlitski:2005pr}. The procedure we have outlined to introduce the axial charge leaves the well-known result for the vector current unaffected. We would like to make a connection between this result and the result for the vector current. If we had not chosen $V(0)=\mu$ and $V(\infty)=0$, but had taken an approach similar to that of the axial field in deriving the vector current, then Eqn.~(\ref{eq:axialcurrent}) would have $a_{31}$ in it instead of $\mu$. We would then compare Eqn.~(\ref{eq:axialcurrent}) with the result from \cite{Metlitski:2005pr} and find that $a_{31}=\mu$, an answer consistent with choosing $V(0)=\mu$ and $V(\infty)=0$. Both methods, either setting the vector potential to zero in the bulk or leaving it free, achieve the same results for the axial current, unlike with the vector current.

If we had instead chosen to move the anomaly to the vector current instead of the axial current, we would still achieve the same results. With the vector current now anomalous, we would be free to define the axial chemical potential as the temporal component of the axial field on the holographic boundary, but we would have to redefine the vector chemical potential to be associated with a constraint in the bulk. 

\section{Discussion}
In the spirit of the bottom-up approach we will now attempt to determine the physical meaning of the boundary condition in Eqn.~\eqref{eq:CME} by comparing it to currents that arise in QCD. In an attempt to model the current we have chosen to make the boundary condition on the horizon nonzero.  We do this because defining an axial chemical potential makes no sense when the axial charge is not conserved. This boundary condition occurs at an IR cutoff we imposed on the bulk. A nonzero boundary condition must be caused by a charge configuration (source) past this boundary further into the bulk (i.e., past the IR cutoff). 

In quantum field theory one can describe the  topological current in a very general way through the introduction of a dynamic theta angle $\theta(\vec{x},t)\neq0$ that is induced by some nonequilibrium process. The existence of these processes in real systems is discussed at length in Section \ref{section:constraints}. The result of this is the induction of a term proportional to
\begin{align}
\sim\theta(\vec{x},t) F_a^{\mu\nu}\widetilde{F}^a_{\mu\nu}.
\end{align}

After a chiral rotation this dynamic theta angle appears as a boundary term. This causes a current to be introduced that is proportional to the derivative of the theta angle. The current that arises from the portion of the current that varies with time is
\begin{align}
\vec{J} = \dot{\theta}(\vec{x},t)\frac{N_c\vec{B}}{2\pi^2}.
\end{align}
Comparing this with Eqn.~(\ref{eq:CME}), we see that it is natural to identify the boundary condition in the bulk with the induced theta angle,
\begin{align}
\label{eq:dictionary}
-A_0(\infty)=\dot{\theta}(\vec{x},t)\,.
\end{align}
In field theory models $\dot{\theta}(\vec{x},t)\neq0$ must be induced by some process. This is the constraint that violates parity and is responsible for a topological current. We have equated this induced $\dot{\theta}(\vec{x},t)$ to the boundary condition that survives Bardeen cancellation. By adding a source beyond the IR we have reproduced the current responsible for the chiral magnetic effect that does not get cancelled due to the Bardeen counter-terms. 

It has been argued that a time-dependent theta angle is equivalent to an axial chemical potential $\dot{\theta} = \mu_5$ \cite{Kharzeev:2009fn}. A change in the theta angle, an external process, induces a chemical potential in the system. A chemical potential must be associated with a conserved charge. This chemical potential must be conjugate to the conserved axial charge of the system. The conserved axial charge then must exist as a configuration in the bulk of the holography. 

We can attempt to see what the charge configuration in the bulk looks like by using the definition for charge $Q = \frac{\partial S}{\partial \mu}$ and assuming that the axial chemical potential is defined as $\mu_5 =  -A_0(\infty)$. The surface terms that contribute are those that contain $A_0$,
\begin{align}
\frac{\partial S}{\partial \mu_5} &=  \frac{\partial }{\partial ( -A_0(\infty))}  \int \!d^4x  \left\{ \frac{1}{4g_s^2}A_0 \frac{e^{-\phi}}{z} \partial_z A_0 \right. \\
&- \frac{N_c}{24\pi^2} \epsilon^{ijk} \left(\frac{2}{6}A_0 V_iV_{jk} + \frac{4}{6} V_i A_0 V_{jk} +  \frac{12}{6} A_0 A_i A_{jk}\right) \\
&  + \left. \left.  \frac{N_c}{24\pi^2}\epsilon^{ijk}  A_0 A_iV_{jk} -  \frac{N_c}{12\pi^2}\epsilon^{ijk}  A_0 V_iV_{jk} \right\} \right|_{z=\infty} \,,
\end{align}
where the first line is from the YM action, the second line is from the CS action, and the last line is from the Bardeen counter-term. Performing the derivative we find that $A_0(\infty)$ is coupled to a charge,
\begin{align}
\label{eq:crazycharge}
Q^5_\text{cutoff} &=  \int \!d^4x \left(- \frac{N_c}{24\pi} \partial_w A_0 + \frac{2N_c}{24\pi^2} \epsilon^{ijk} A_iA_{jk} \right. \\ & \left.\left.+  \frac{3N_c}{24\pi^2} \epsilon^{ijk} V_iV_{jk} - \frac{N_c}{24\pi^2} \epsilon^{ijk} A_iV_{jk} \right) \right|_{z=\infty}\,.
\end{align}
Let us look closer at the form of this charge. 

One would expect the axial charge of the system to be given by integrating over the zeroth component of the axial current given by \eqref{eq:currentequations2} and \eqref{eq:currentequations4},
\begin{align}
Q^5 &=   \int \!d^4x  J_{0}^5\,,\\
\label{eq:regularcharge}
& = \int \!d^4x \left( - \frac{N_c}{24\pi} \partial_w A_0 + \frac{2N_c}{24\pi^2} \epsilon^{ijk} A_iA_{jk} \right).
\end{align}
We see that the charge derived by assuming the boundary condition at infinity is an axial chemical potential \eqref{eq:crazycharge} differs from the charge of the system found by looking at the zeroth component of the axial current \eqref{eq:regularcharge}. In fact, if we set $A_i=0$, as Rubakov does, the two charges, $Q^5$ and $Q^5_\text{cutoff}$, differ by exactly the amount Rubakov used to define his conserved charge given by equation \eqref{eq:rubakovcharge},
\begin{align}
Q^5_\text{Rubakov} =   \int \!d^4x  \left( J_{0}^5 +  \frac{3N_c}{24\pi^2} \epsilon^{ijk} V_iV_{jk}\right).
\end{align}
The difference is that the charge we calculated for the boundary condition is evaluated at $z=\infty$, while Rubakov's rests on the holographic boundary. That the boundary value of the axial field  $-A_0(\infty)$ is coupled to an object that matches the form of Rubakov's charge is evidence that $-A_0(\infty)$ is related to the axial chemical potential. What configuration past the IR would cause this boundary condition is unknown. 

A similar definition of the axial chemical potential was necessitated in the work of \cite{Gynther:2010ed}\footnote{Thank you to Karl Landsteiner for pointing this out to us.}, working in the context of linear response theory, where the horizon of an AdS-Schwarzschild black hole provides the screening of the IR physics. They define, based on the work of \cite{Ghoroku:2007re}, the chemical potential as the difference in energy between the boundary  and bulk. They then use the Kubo formula to derive their currents. A boundary condition from the bulk affects the current on the boundary.

\vspace{-0.2cm}
\begin{acknowledgments}
\vspace{-0.3cm}
We would like to thank Ariel Zhitnitsky, Moshe Rozali, and Dam Thanh Son for guidance and many useful discussions. Thank you to Karl Landsteiner for pointing out Ref. \cite{Gynther:2010ed} to us. This research was supported in part by the National Science and Engineering Research Council of Canada. 
\end{acknowledgments}


%

\end{document}